\def\lj{{\lambda_J}}
\def\bJ{{\beta_{_{J}}}}

\documentclass[groupedaddress,showkeys,superscriptaddress,11pt,a4paper]{article}
\usepackage {amsmath, amssymb}
\usepackage{color}
\usepackage[T1]{fontenc}
\usepackage[utf8]{inputenc}
\usepackage{authblk}
\usepackage{graphicx}
\usepackage{epsfig}
\usepackage[small,bf]{caption}
\usepackage{subcaption}

\usepackage{fullpage}

\begin{document}

\bibliographystyle{abbrv}

\title{Sine-Gordon breathers generation in driven long Josephson junctions}

\author[1,2]{Claudio Guarcello\thanks{e-mail: claudio.guarcello@unipa.it}}
\author[1]{Davide Valenti}%\thanks{e-mail: davide.valenti@unipa.it}}
\author[1,2,3]{Bernardo Spagnolo\thanks{e-mail: bernardo.spagnolo@unipa.it}}
\affil[1]{Group of Interdisciplinary Physics Universit\`a di Palermo and CNISM, Unit\`a di Palermo Viale delle Scienze, Edificio 18, 90128 Palermo, Italy}
\affil[2]{Radiophysics Department, Lobachevsky State University, 23 Gagarin Avenue, 603950 Nizhniy Novgorod, Russia}
\affil[3]{Istituto Nazionale di Fisica Nucleare, Sezione di Catania, Via S. Sofia 64, I-90123 Catania, Italy}

\renewcommand\Authands{ and }
\date{}
\maketitle

\begin{abstract}

We consider a long Josephson junction excited by a suitable external ac-signal, in order to generate control and detect breathers. Studying the nonlinear supratransmission phenomenon in a nonlinear sine-Gordon chain sinusoidally driven, Geniet and Leon~\cite{Gen02,Gen03} explored the bifurcation of the energy transmitted into the chain and calculated a threshold $A (\omega)$ for the external driving signal amplitude, at which the energy flows into the system by breathers modes. I numerically study the continuous sine-Gordon model, describing the dynamics of the phase difference in a long Josephson junction, in order to deeply investigate the ``continuous limit'' modifications to this threshold. Wherever the energy flows into the system due to the nonlinear supratransmission, a peculiar breather localization areas appear in a $(A, \omega)$ parameters space. The emergence of these areas depends on the damping parameter value, the bias current, and the waveform of driving external signal. The robustness of generated breathers is checked by introducing into the model a thermal noise source to mimic the environmental fluctuations. Presented results allows one to consider a cryogenic experiment for creation and detection of Josephson breathers.

\end{abstract}

The behavior of a nonlinear system excited by an external sinusoidal driving and the bifurcation of the energy transmission into the system were deeply studied by Geniet and Leon~\cite{Gen02,Gen03}. The Floquet theorem~\cite{Bri46} states that, for linear waves in a periodic structure, a forbidden gap of frequency exists. The frequency gap between acoustic and optical branches in the vibration modes of a diatomic chain well embodies this picture. Waves with frequency within this forbidden range esponentially vanish in the medium (evanescent waves). In nonlinear systems excited by plane waves, the appearance of gap solitons~\cite{Che87} and the localization of energy due to nonlinear instability are matter of many systems~\cite{Ost03, Des03, Lou03, Joh90, Ess06, Sca94}. If the nonlinear medium, characterized by a forbidden band gap (FBG), is irradiated to one end by a sinusoidal drive, with frequency within the gap, the energy transmission can be supported. This phenomenon is called \emph{nonlinear supratransmission} (NST) and is nowadays reported in several other contexts, from Bragg media~\cite{LeoSpi04, Tav98}, coupled-wave-guide arrays (nonlinear Schr\"{o}dinger model)~\cite{Leo03, Leo04}, optical waves guide arrays~\cite{Kho04}, Fermi-Pasta-Ulam model~\cite{KhoLep04}, Klein-Gordon (KG) electronic network~\cite{Bod10}, chains of coupled oscillators~\cite{Mac07}, discrete inductance-capacitance electrical line~\cite{Koo07} to generic multicomponent nonintegrable nonlinear systems~\cite{Ang10}. The manifestation of NST in both SG and KG equations, points out that this process is not dependent on the integrability, but is a feature of systems possessing a natural forbidden band gap. Geniet and Leon found a very simple rule proving an explicit formula for the bifurcation of the energy in a $(\omega, A)$ diagram, called the \emph{bifurcation diagram}, where $\omega$ and $A$ are the frequency and the amplitude of the external signal, respectively. The energy flows into the medium as soon as the amplitue $A$ of the driving, at frequency $\omega$, exceeds the maximum amplitude of the static breather of the same frequency. This energy then travels through the system by means of plasma waves and nonlinear localized excitations, that is solitons, antisolitons and breathers. Geniet and Leon extended the results, obtained for a SG chain, to its continuous version, used to describe a long Josephson junction (LJJ) whose extremity is subjected to a sinusoidal excitation (\emph{Neumann boundary condition}).
The soliton solutions are strongly stable, maintaining their shape after collisions or reflections, and producing clean evidences in the I-V characteristics of the junction. A breather is another travelling SG solution formed by a soliton-antisoliton bounded couple, oscillating in an internal frame with a proper internal frequency. The magnetic flux associated with a breather is zero, as the voltage difference across the junction. Indeed, considering that the time variations of $\varphi$ produce a voltage difference across the junction according to the a.c. Josephson relation, the rapid oscillations of the breather phase result in a mean voltage close to zero (or beyond the sensitivity of nowadays high frequencies oscilloscopes). Other pratical difficulties exist: i) a breather has to be efficently generated and trapped, for a sufficent long time, in a confined area to allow measurements, ii) it decays esponentially in time, even more so considering the damping affecting the system, and iii) an applied bias current tends to split the breather in a soliton/antisoliton couple. These are the reasons that underlie the lack of experimental evidences and confirmations of breathers in LJJ. Observations of breathers, called \emph{rotobreathers}, in Josephson junctions (JJ) are reported only using ``Josephson ladders'', i.e. a ladder composed by small JJs~\cite{Bin00, Bin02, Ust01, Sch03,Seg14}. Our investigation points towards an experimental setup devoted to the exclusive generation and detection of breathers in LJJ. The influence of the environment should be take into account to well describe a real experimental framework. Bodo \emph{et al}~\cite{Bod09, Bod10, Bod13} enhanced the deterministic analisys developed by Geniet and Leon in a discrete SG chain, using a sinusoidal excitation corrupted by an additive white Gaussian noise. Considering driving amplitudes below the critical value, they found that noise induced breathers can be generated with a probability depending on the noise intensity. 

To check the robustness of the generated breathers, we add to the perturbed SG model a Gaussian white source, to mimic the environment effect on the breather's dynamics. We calculate the percentage of surviving breather, including a white noise source of given amplitude. \\

\emph{The continuous SG model}. $-$ The electrodynamics of a LJJ in the geometrical configuration of overlap, is described in terms of a nonlinear partial differential equation for the order parameter $\varphi$, that is the sine-Gordon equation~\cite{Bar82,Lik86}. 
Here $\varphi$ is the phase difference between the wave functions describing the superconducting condensate in the two electrodes. Our analysis includes a quasiparticle tunneling term and an additional stochastic contribution, $i_{TN}(x,t)$, representing the Gaussian thermal noise effects. However, the surface resistance of the superconductors is neglected. The resulting \emph{perturbed} SG equation reads
\vskip-0.2cm
\begin{eqnarray}
\varphi_{tt}(x,t)+\bJ \varphi_{t}(x,t)-\varphi_{xx}(x,t) = i_b(x,t)-\sin(\varphi(x,t))+i_{TN}(x,t),
\label{sine_Gordon}
\end{eqnarray}
where a simplified notation has been used, with the subscript of $\varphi$ indicating the partial derivative in that variable. This notation will be used throughout all paper. The SG equation is written in terms of the dimensionless $x$ and $t$ variables, that are the space and time coordinates normalized to the Josephson penetration depth $\lj$ and to the inverse of the plasma frequency $\omega_p$ of the junction, respectively. In the limit of small amplitude oscillations, the JJ plasma frequency corresponds to the oscillation frequency in the bottom of the a washboard potential well, modified by the presence of a bias current~\cite{Bar82}.
Damping parameter is $\bJ=(\omega_p R_NC)^{-1}$, where $R_N$ and $C$ are the effective normal resistance and capacitance of the junction. 
The terms $i_b(x,t)$ and $sin(\varphi)$ of Eq.~(\ref{sine_Gordon}) are the bias current and supercurrent, respectively, both normalized to the JJ critical current $i_c$. Eq.~(\ref{sine_Gordon}) is solved imposing the following boundary conditions
\begin{equation}
\varphi_{x}(0,t) = f(t) \qquad\qquad \varphi_{x}(L,t) = 0,
\label{boundaries}
\end{equation}
where $f(t)$ is the external driving pulse used to excite the LJJ and generate the breathers.

The unperturbed SG equation is
\begin{equation}
\varphi_{xx}(x,t)-\varphi_{tt}(x,t)=\sin(\varphi(x,t)).
\label{SGunperturbed}
\end{equation}
and admits solutions in the traveling wave form $f=\varphi(x-ut)$~\cite{Bar82}, 
\begin{equation}
\varphi(x,t)=4\arctan \left \{ \exp \left [ \pm \frac{\left(x-ut \right )}{\sqrt{1-u^2}} \right ] \right \}, 
\label{SGkink}
\end{equation}
where $u$ is the wave propagation velocity normalized to the speed of light, and is called \emph{Swihart velocity}. Eq.~(\ref{SGkink}) represents a single \emph{kink}, or \emph{soliton}. The signs $+$ and $-$ indicate a $2\pi$-kink (soliton) and a $2\pi$-antikink (antisoliton), respectively. In this framework, $\varphi$ gives a normalized measure of the magnetic flux through the junction, so that Eq.~(\ref{SGunperturbed}) can also represent the motion of a single fluxon (or antifluxon). In fact, starting from simple electrodynamic considerations~\cite{Bar82}, it is possible to obtain a simple relation between the magnetic field $H(y)$ and the spatial derivative of the phase difference
\begin{equation}
\varphi_x=2\pi\frac{d H(y)}{\Phi_0}, \label{phiX-H}\qquad\text{integrating over the entire JJ lenght:} \qquad \varphi(L)-\varphi(0)=2\pi\frac{\Phi_H}{\Phi_0}
\end{equation}
where $\Phi_0=hc/2e$ is the fluxon and $d=\lambda_L+\lambda_R+t$ is the magnetic penetration, $\lambda_L$ and $\lambda_R$ are the London depths in the left and right superconductors and $t$ is the interlayer thickness. If the phase evolution shows a single $2\pi$-kink, a single fluxon will propagate along the junction. Analytic expressions of travelling-wave solutions of the unperturbed SG equation can be explicitly written also for N-solitons, for kink-antikink collisions and for breathers. Also the damped and biased SG equation has slightly perturbed kink as solutions. Indeed, after a series of trasformations in zeroth order in bias and damping parameter, the PDE describing the evolution of this perturbed system reduces to the unperturbed SG differential equation~\cite{Iva13}. 

The analytic expression of a \emph{stationary breather} oscillating with frequency $\omega$ is:
\begin{equation}\label{StatBreather_Eq}
\varphi(x,t)=4\arctan \left \{ \frac{\sqrt{1-\omega ^{2}}}{\omega } \frac{\sin \left ( \omega t \right )}{\cosh \left ( x\sqrt{1-\omega ^{2}} \right )}\right \}.
\end{equation}
The SG solutions are invariant under Lorentz transformations, so this stationary breather can be boosted into a moving frame, resulting in the following \emph{moving breather} travelling with an envelope velocity $v<1$,
\begin{equation}\label{DynBreather_Eq}
\varphi(x,t)=4\arctan \left \{ \frac{\sqrt{1-\omega ^{2}}}{\omega } \frac{\sin \left [ \gamma\omega \left ( t-vx \right ) \right ] }{\cosh \left [ \gamma\sqrt{1-\omega ^{2}} \left (t \right )\left ( x-vt \right ) \right ]}\right \},
\end{equation}
where $\gamma=\left ( 1-v ^{2}\right )^{-1/2}$ is the Lorentz factor.\\ \\

\emph{The nonlinear supratransmission: from discrete to continuous SG}. $-$ Studying a discrete SG chain of $N$ damped oscillators:
\begin{equation}\label{DiscrSG}
\ddot{u}_n-c^2(u_{n+1}-2u_{n}+u_{n-1})+\sin u_n=-\gamma(n)\dot{u}_n \qquad n=1,...,N
\end{equation}
with boundary conditions
\begin{equation}\label{DiscrSG_bc}
u_0(t)=A\sin (\omega t) \qquad\qquad u_n(0)=0 \qquad\qquad \dot{u}_n(0)=0,
\end{equation} 
where $c$ is the coupling factor and $\gamma$ the damping coefficent, Geniet and Leon~\cite{Gen02,Gen03} deduced, exciting the JJ in the frequencies forbidden gap ($\omega<1$), a threshold $A_s(\omega)$:
\begin{equation}\label{DiscrSG_bc}
A_s(\omega)=4\arctan \left [ \frac{c}{\omega} \textup{arccosh}\left ( 1+\frac{1-\omega^2}{\omega} \right ) \right ],
\end{equation} 
above which NST accours and the system permits the energy trasmission by means of nonlinear modes generation (breathers, kinks, antikinks). 
\begin{figure}[t]
      
        \begin{subfigure}{0.51\textwidth}
                \includegraphics[width=80mm]{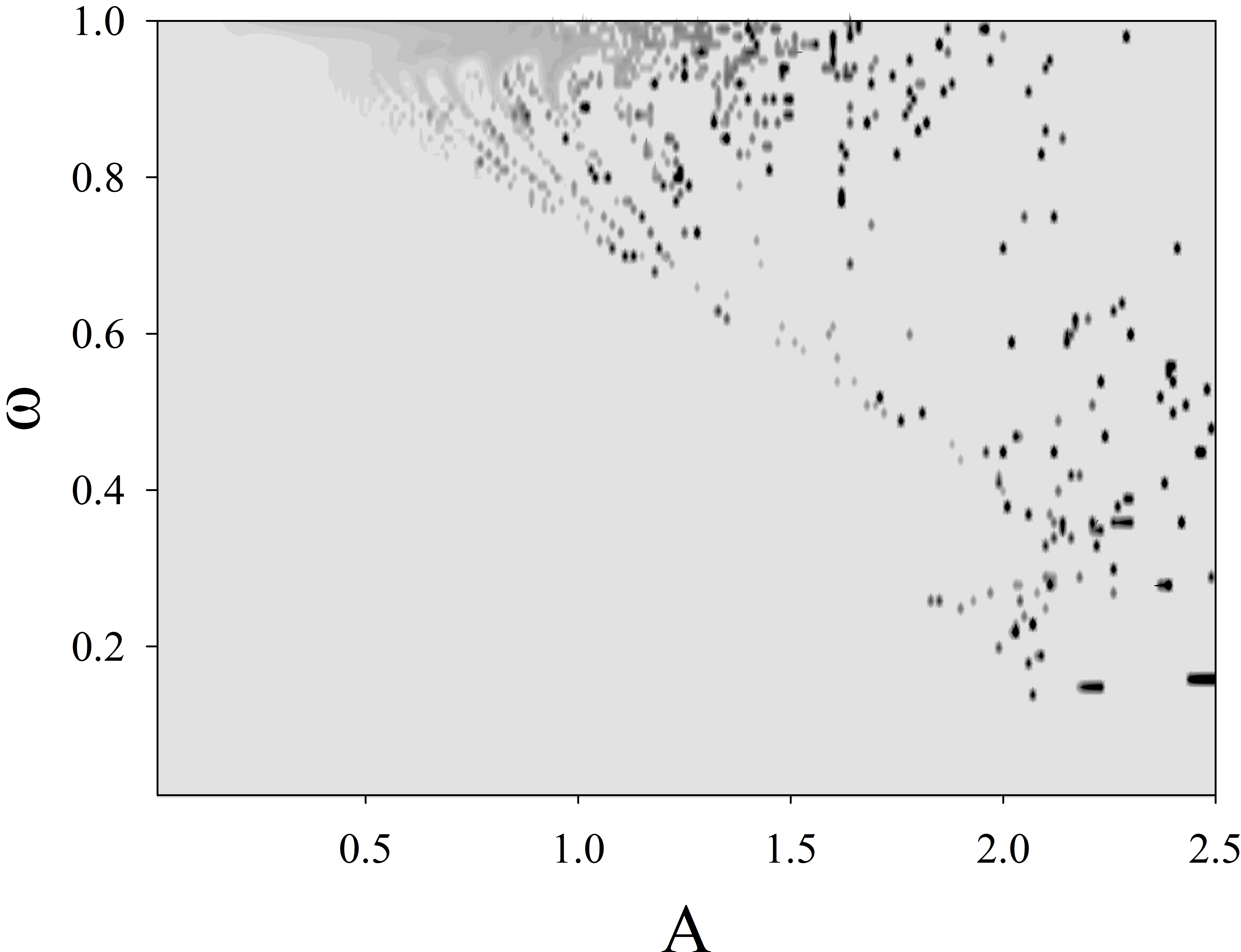}
                \caption{}
                \label{Results(a)}
        \end{subfigure}%
        \begin{subfigure}{0.60\textwidth}
                \includegraphics[width=80mm]{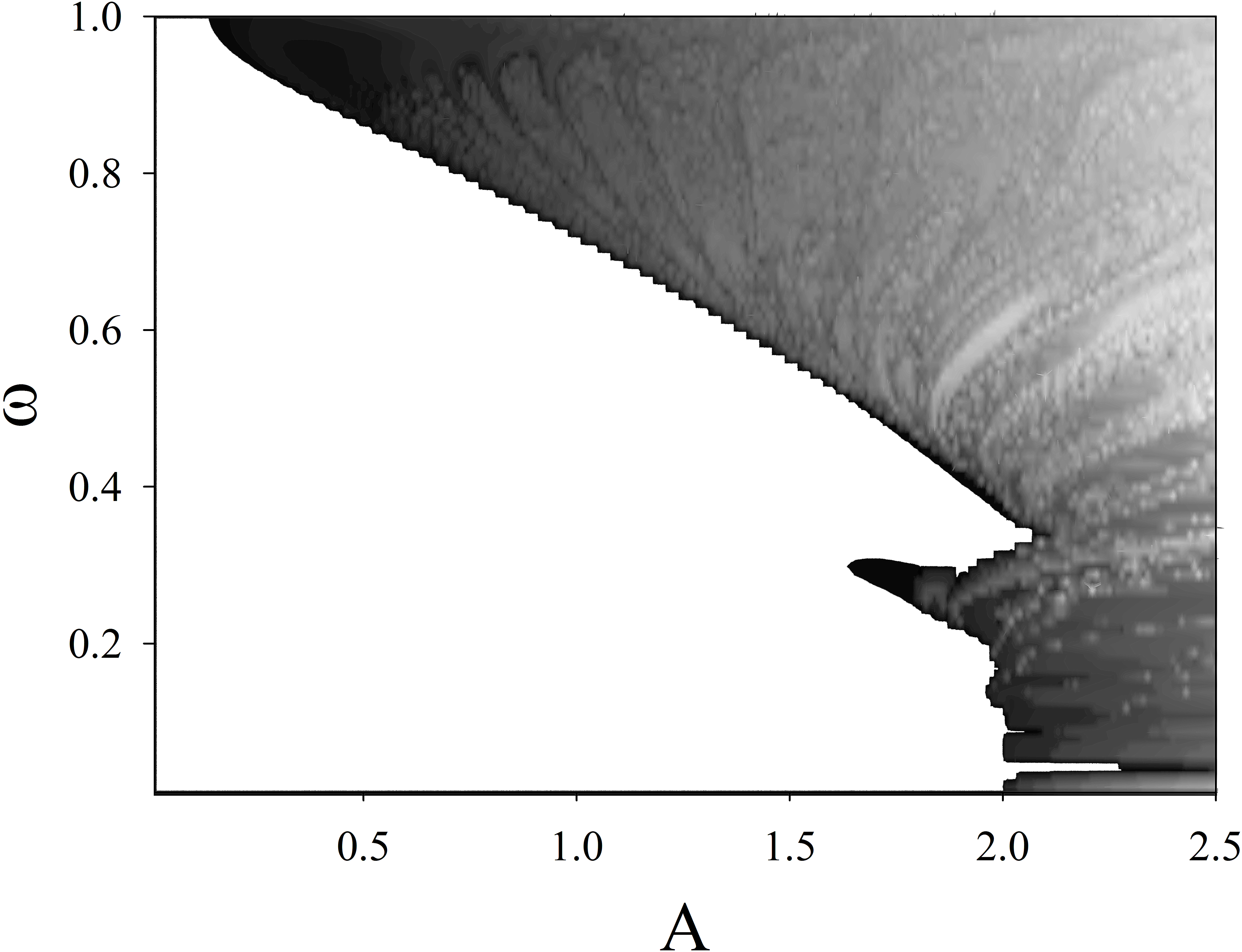}
                \caption{}
                \label{Results(b)}
        \end{subfigure}%
        \caption{a) Breathers-branches localized in a $(A,\omega)$ parametric 2D-space. b) Energy injected into the medium in a $(A,\omega)$ parametric 2D-space.}\label{Results}
\end{figure}
In the continuous limit, that is $1/c\rightarrow0$, using the boundary condition:
\begin{equation}\label{ContinSG_bc}
{\partial _{x}u|}_{x=0}=B\sin (\omega t).
\end{equation} 
the resulting threshold is:
\begin{equation}\label{ContinSG_threshold}
B_s=2(1-\omega^2).
\end{equation} 
The $B_s$ expression is calculated imposing that the system adapts its phase profile to the breather derivative at the boundary ${\partial _{x}u |} _{x=0}={\partial _{x}u_b |} _{x=0}$. Varying the position of the breather's center to maximize ${\partial _{x}{u_b}|}_{x=0}$, the simple expression for the maximum amplitude, see Eq.~\ref{ContinSG_threshold}, results.\\

\emph{Numerical results}. $-$ The deterministic analysis is carried on setting $i_{TN}(x,t)=0$ in the Eq.~\ref{sine_Gordon}. Two different bifurcation diagrams are shown in Fig.~\ref{Results}. Due to the time normalization performed for the SG model, set $\omega<1$ means explore the frequencies lower than the JJ plasma one. In the $(A,\omega)$ parametric space shown in Fig.~\ref{Results(a)}, the dark spots indicate $(A,\omega)$ values in correspondence of which only breathers are generated. In particular, the breathers tend to localize in \emph{branches} above the bifurcation threshold, see Eq.~\ref{ContinSG_threshold}, whereas are totally absent below the threshold. The grey background of Fig.~\ref{Results(a)} indicates the absence of breathers modes uniquely generated into the junction. In detail, for low $(A,\omega)$ values (below the threshold $B_s$) there is no NST effect, while above the threshold every kind of nonlinear excitation can propagate through the medium. Plasma waves, generated by the external signal and resulting from the breathers decay, are also detected with the breathers. Beetwen these breathers-branches, every combination of kinks, antikinks, breathers and plasma waves can be detected, but due to the stable nature of solitons and the damping of the medium, after a short time only soliton solutions survive into the system. The branches for high amplitude values, that is $A\gtrsim 1$, corresponde to multiple generation of stationary and moving breathers. The results for low frequencies ($\sim 0.2$) and high amplitudes are consistent with Geniet and Leon's results~\cite{Gen03}. Changing the bias current value, the damping parameter and the duration of the external signal, modifications in the breathers arrangement come to light. 

The analisys of the energy injected into the system is displayed in Fig.~\ref{Results(b)}. This $(A,\omega)$ bifurcation diagram shows the same branched structures, suggesting moreover that the breathers tends to localize into isoenergetic regions of this diagram, in particular where the energy is slightly lower than the sorrounding areas. Indeed, the energy of a breather~\cite{Sco04}:
\begin{equation}\label{Br_Energies}
 E_{\mathrm{breather}}=16\gamma(1-\omega^2)
\end{equation} 
is less then the energy of a kink/antikink couple
\begin{equation}\label{Kink_Energies}
E_{\mathrm{kink}}=E_{\mathrm{antikink}}=8\gamma .
\end{equation} 
The Fig.~\ref{Results(b)} shows also a large white area, sign of the lack of NST (no energy flows into the system).

\begin{figure*}[t]
\centering
\includegraphics[width=130mm]{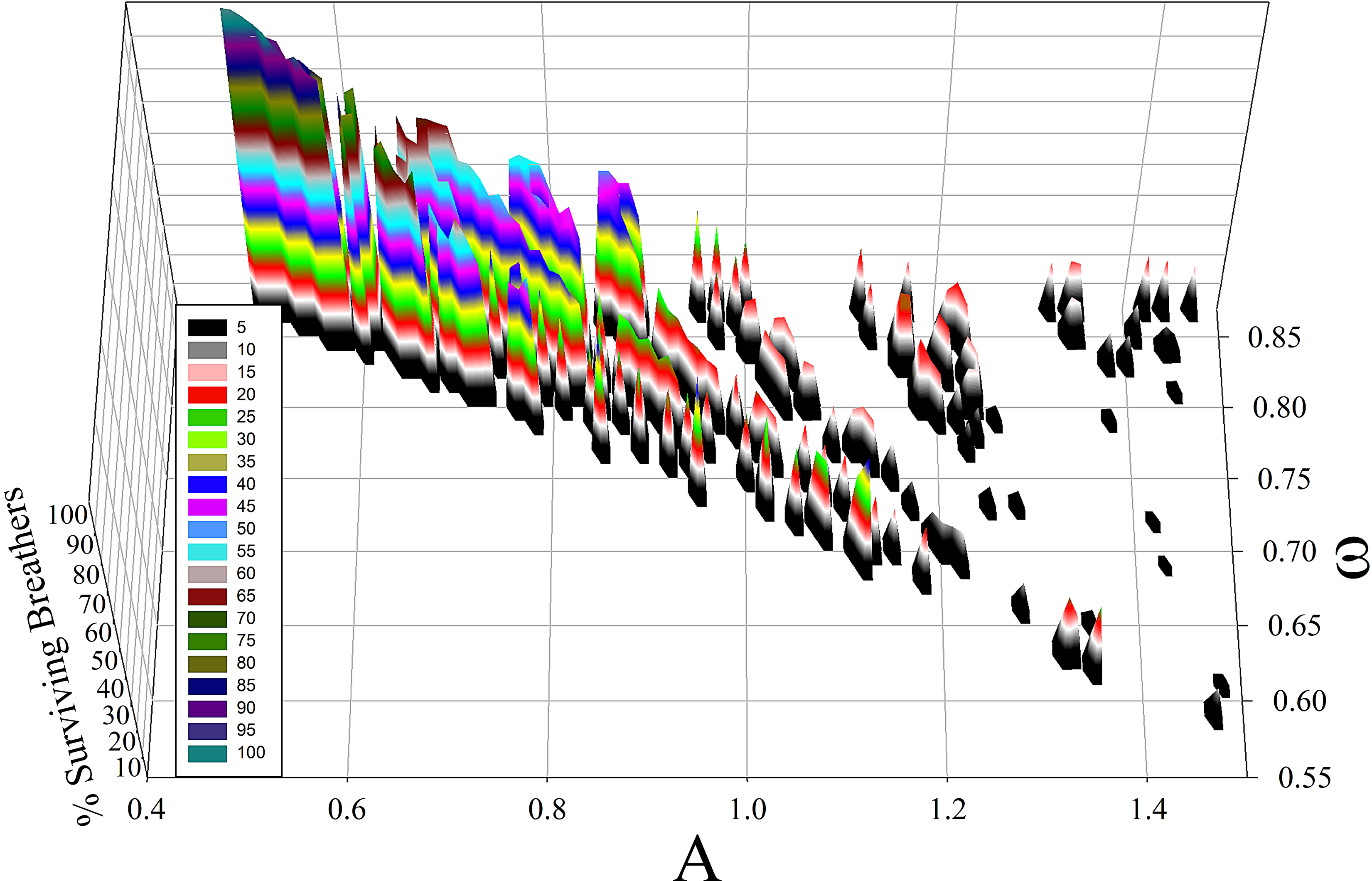}
\caption{(Color online) Percentage of surviving breathers as a function of amplitude $A$ and frequency $\omega$ of the external drive, setting $\gamma_{TN}=10^{-3}$ and performing $N=10^3$ numerical realizations.}
\label{Breathers_Energy}
\end{figure*}

To check the robustness of the generated breathers, a nonzero noise amplitude is setted. The stochastic investigation is developed performing $N=10^3$ numerical realizations, setting the $(A,\omega)$ values that deterministically corresponde to breathers generation, including into the SG model, see Eq.~\ref{ContinSG_threshold}, a white noise contribute with amplitude $\gamma_{TN}=10^{-3}$. The thermal fluctuations can supply the system with an amount of energy large enough to create kinks instead of a breather, or to break a generated breather into a kink-antikink couple. The results in the Fig.~\ref{Breathers_Energy} shows the percentage of surviving breathers as a function of the amplitude $A$ and the frequency $\omega$. For large $A$ values, little amount of thermal energy can be enough to destroy a breather. Anyway, a suitable $(A,\omega)$ values range exists, in which breathers are still generated by external excitation with high probability despite of the thermal influence. \\

\emph{Conclusions}. $-$ We explore the breathers generation and propagation in a long Josephson junction externally irradiated by a suitable excitation, changing the amplitude $A$ and the frequency $\omega$ of the signal. The analysis is computationally developed in the framework of the damped and biased continuous sine-Gordon equation, inserting a Gaussian white noise source to include the environmental influence. Taking a cue from the Geniet and Leon results about the nonlinear supratransmission in a discrete sine-Gordon chain, we discover a new peculiar localization of the breathers in branches in a particular $(A,\omega)$ diagram. We distinguish two parameters regions: for low values of driving frequencies and amplitudes, no energy flowing into the system is detected, otherwise every kind of nonlinear excitations travels through the junction. When this occurs, we are able to recognize where only breathers are generated. The deterministic analysis is improved including a white noise contribute, to evaluate the percentage of breathers enduring the thermal effects.\\

\emph{Acknowledgements}. $-$ Authors acknowledge the financial support of Ministry of Education, University, and Research of Italian Government (MIUR). One of the authors, C G, strongly thanks the KIT ``Karlsruhe Institute of Technology'' for the hospitality at the laboratory of Professor A. Ustinov in the spring of 2013, where the original idea of this work has been developed.

%\bibliography{biblio}

\end{document}